# Optical Properties of Nanoporous Al₂O₃ Matrices with Ammonium Dihydrogen Phosphate Crystals in Nanopores

Nazariy Andrushchak, Dmytro Vynnyk, Anatoliy Andrushchak
Lviv Polytechnic National University
Lviv, Ukraine
nandrush@gmail.com

Volodymyr Haiduchok
Scientific research company "Electron-Carat" - branch of the private joint stock company "Concern-Electron"
Lviv, Ukraine

Yaroslav Zhydachevskyy, Markiyan Kushlyk
Institute of Physics, Polish Academy of Sciences
Warsaw, Poland

*Abstract*— In this paper, the technology of growing ammonium dihydrogen phosphate (ADP) crystals in pores of nanoporous aluminum oxide ($Al_2O_3$) matrices was presented. On the grown nanocomposites, the optical properties of such structures were studied. In particular, the direct and diffuse optical transmission, as well as specular reflection of Al2O3 matrices with ADP crystals grown in nanopores, were analyzed. From the performed experimental studies, it can be concluded that the transmission and scattering of pure $Al_2O_3$ porous matrices and with ADP crystals in nanopores depends largely on the properties of the $Al_2O_3$ matrices and the technology of their production.

## I. Introduction

In recent years, different methods have been widely applied to obtain nanoobjects of various materials by introducing into their empty nanoscale channels of dielectric matrices of opals, porous glasses, asbestos and zeolites [1].

An important issue is the creation of new functional materials with adjustable optical properties used in membrane technologies, optics, quantum electronics and analytical instrumentation. Prospective matrices for such materials are porous glass due to their unique characteristics such as adsorption, diffusion, optical and other characteristics in combination with adjustable structural parameters of the nanosized range.

The study of composite materials based on bulk or thin film porous matrices (submicron or nanometer pores of which are filled with ferroelectric materials), is of a great scientific and practical interests related to the intensive development of nanotechnologies [2-4].

When filling the pores of the porous matrix with substances that have active physical properties, the functional capabilities of the introduced components are significantly expanded and the practical significance of such structures increases substantially. Therefore, nanostructured composite materials based on porous oxides [6] are under the intensive research, where aluminum oxide ($Al_2O_3$) stands out by the comparative simplicity of its obtaining by sulfuric, oxalic and phosphoric acids in electrolytes [6]. The self-organized array of pores that were formed during the digestion differ in the uniform density $\sim 10^9$-$10^{10}$ cm$^{-2}$ with an average pore size.

Ammonium dihydrogen phosphate (ADP) is one of the first crystals found to be practical, mainly due to its nonlinear optical and electro-optical characteristics. The ADP crystal is a negative monochromatic crystal belonging to the tetragonal symmetry group (point group *42m*) with parameters of the lattice: $a$=7,495Å, $c$=7,548Å [7]. The ADP crystal is characterized by wide optical opacity of about 0.184–1.3 μm with a width of the band gap $Eg$ = 6.8 eV [8]. The experimental value of the angle of phase-synchronization of the second optical harmonic generation process of the laser radiation with a wavelength of 1064 nm ($o + o \rightarrow e$) at room temperature is: $\Theta$ = 41,9° [9], $\Theta$ = 42° [10]. Therefore, the study of $Al_2O_3$ matrices with ADP crystals in nanopores is of considerable scientific interest.

In [11-16] we have developed methodologies and specific setups for experimental testings of the crystalline materials in a wide frequency range. However, with development of new materials new methods for fast evaluation of such materials are crucial and become even more needed if pores of the composite are filled with another crystalline material [17], which can be used for creating of new optical cells with increased efficiency [18]; thus, new phenomena can be investigated [19,20]. For the preliminary analysis of new crystalline structures, numerical methods for evaluating physical properties (e.g. refractive index) are used [21].

## II. Growing of Nanocrystals

In a chosen nanoporous matrix of $Al_2O_3$, the ADP nanocrystals inside the pores were grown from a saturated solution of ADP. To prepare a saturated solution of ADP, pre-prepared crystals that were dissolved in distilled deionized water were used to grow nanocrystallites in pores of the $Al_2O_3$ porous matrix.

The nanoporous plates before the start of operation undergo so-called activation in a vacuum oven at 250°C and rinsing with dry nitrogen at a temperature of 150 - 200°C. After these activations weighing of plates was performed. In order to control the "weight gain", the weighing process was carried out before and after the growth of nanocrystals in a nanoporous matrix.

Using the described above process of activation the nanoporous $Al_2O_3$ matrices with different pore diameter were prepared. The thickness of each sample is $d = 0.1$ mm and the dimensions $S = 10\times10$ mm$^2$. These samples were kept for 3-5 hours at 56.5°C and then soaked into the saturated solution of ADP in specially prepared test tubes (dents) covered from the one side with the nylon mesh on which the samples were places (see Fig. 1).

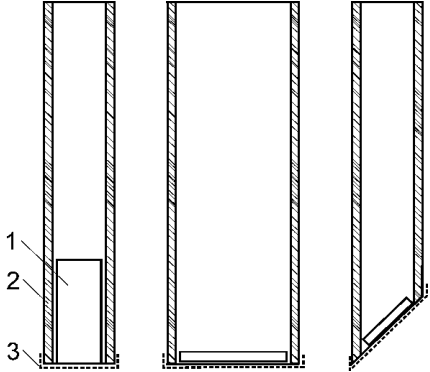

Fig. 1. Glass tubes for carrying out growth manipulations with porous $Al_2O_3$ matrices in a thermostat: 1 – porous glass; 2 – glass tube; 3 – nylon mesh

The nanoporous plates were kept in the saturated solution of ADP for 2.5 hours, then removed from the solution, and wiped first with a little damp and later with a dry cotton cloth. After that, the samples were dried at 50°C for 20 hours. The growth process of the nanocrystals was carried out in the so-called "dry" thermostat – a drying cabinet with a temperature control and the accuracy of ±0.5°C. After completion of all growth procedures, samples were placed in a silica gel desiccator (moisture absorber).

Re-weighing of the nanoporous $Al_2O_3$ matrices filled with the ADP crystals showed the increase in weight of 0.0203g. For all growing of ADP nanocrystals in the pores of the $Al_2O_3$ matrix that are presented in the current paper, we were using $Al_2O_3$ matrices from SmartMembranes Company (Germany).

### III. EXPERIMENTAL RESULTS

Before the growth of nanocrystals in pores of $Al_2O_3$, we have taken the transmission spectra of these matrices of the 100 μm thickness and the diameter of nanopores equal to 25 nm, 35 nm, 60 nm and 80 nm. The transmission spectra were taken on a Shimadzu UV-3600 spectrophotometer (Japan).

The choice of nanopore diameter in the $Al_2O_3$ matrix for further grow of ADP nanocrystals were done based on the analyzed results from Fig. 2. As it can be seen from the direct transmission spectra measured normally (Fig. 2), the magnitude of the direct transmittance of such matrices depends largely on the diameter of nanopores and the larger the entire measured wavelength range, the smaller the size of the nanopores. Thus, for further experimental investigations and growth of ADP crystals inside these pores the samples with the pore diameter of 25 nm were chosen. These measurements are in good agreement with the results of work in [22].

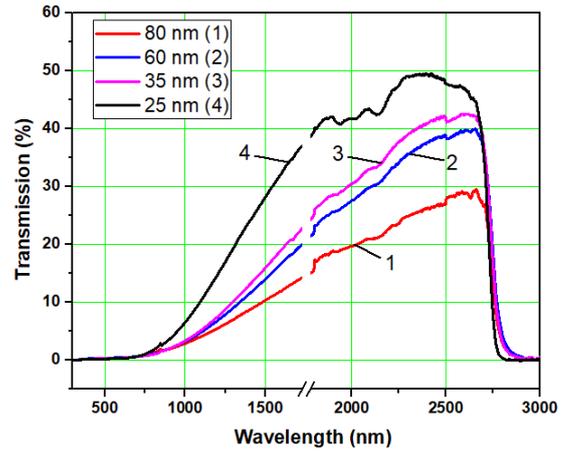

Fig. 2. Transmission spectra of $Al_2O_3$ matrices with different diameter of nanopores: 1 - diameter of nanopores - 80 nm; 2 - diameter of nanopores - 60 nm; 3 - diameter of nanopores - 35 nm; 4 - diameter of nanopores - 25 nm

However, as it was discovered experimentally, among the $Al_2O_3$ matrices with a thickness of 100 μm and a diameter of nanopores 25 nm, there were some samples (#5 and #6), where the direct transmission spectra are different from those that are shown in Fig. 2. The direct transmission spectra of such samples are shown in Fig. 3. As can be seen from Fig. 3, the transmission spectra of these matrices are largely different from the spectra of the matrices $Al_2O_3$ depicted in Fig. 1 in the short-wave region of the spectrum.

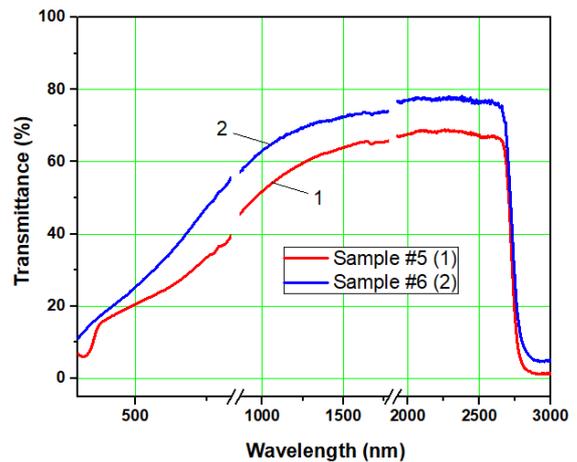

Fig. 3. Transmission spectra of $Al_2O_3$ matrices (#5 and #6) with a thickness of 100 μm and a diameter of pores 25 nm

In Fig. 4 and Fig. 5 the transmission spectra of $Al_2O_3$ matrices with ADP nanocrystals grown in nanopores from a water solution with ADP crystals are shown.

As can be seen from Fig. 4 the growth of nanocrystals in nanopores with a diameter of 80 nm practically does not change the nature of the transmission spectrum of this matrix in comparison with a pure $Al_2O_3$ matrix with a pore diameter of 80 nm (see Fig. 2).

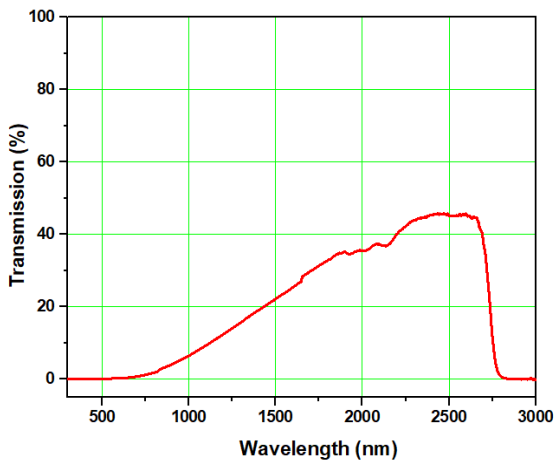

Fig. 4. The transmission spectrum of $Al_2O_3$ matrices with ADP crystals grown in nanopores with the diameter of pores 80 nm

However, as it can be observed from the transmission spectra of $Al_2O_3$ matrices with a pore diameter of 25 nm, with ADP crystals grown in nanopores (Fig. 5), there is a significant difference between the spectra of samples #5 and #6. The character of the transmission spectrum of sample #6 is the same as the one that is shown in Fig. 4, while the transmission of sample #5 in the investigated range of wavelengths (0.3-3.0 microns) does not exceed 0.5%. Thus, as it can be seen that the sample practically does not pass light throughout the whole investigated range.

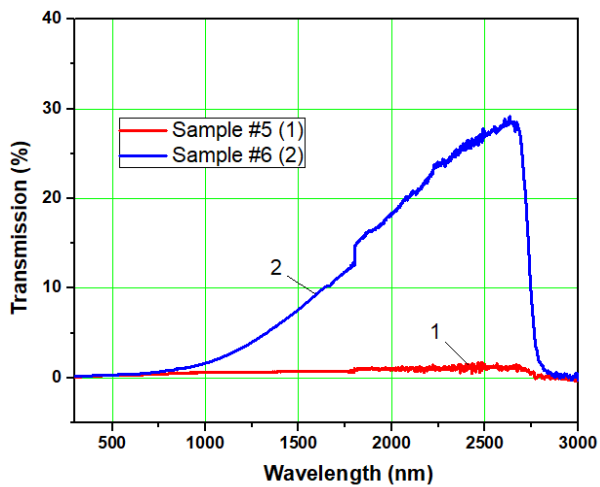

Fig. 5. Transmission spectra of $Al_2O_3$ matrices with ADP crystals grown in nanopores with the diameter of pores Ø25 nm

The specular reflection spectra of all samples with pure nanopores of different sizes and with nanopores, in which the ADP crystals were grown, are practically identical and differ only in the magnitude of the reflection coefficient in the visible spectral range. The specular reflection spectrum of the $Al_2O_3$ matrix with ADP crystals grown in nanopores with a diameter of 25 nm is presented in Fig. 6.

Since the $Al_2O_3$ matrices with nanopores have significant light scattering (see Fig. 7), the transmission spectra of specimens #5 and #6 were taken including both the direct and diffuse transmittance (see Fig. 8), as well as the diffuse transmittance only (Fig. 9). These spectra were measured using a Varian Cary 5000 spectrophotometer with external diffuse reflectance accessory DRA-2500.

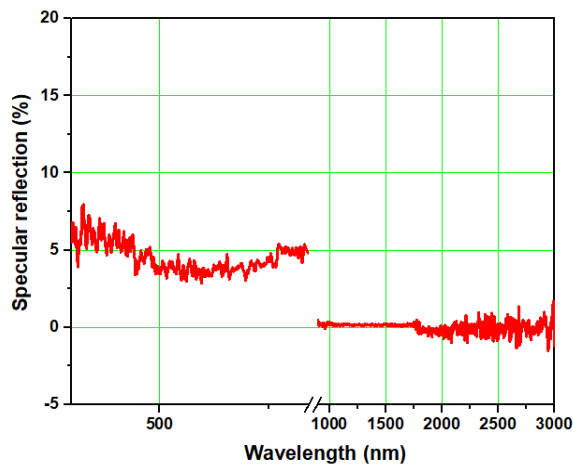

Fig. 6. Specular reflection spectra of $Al_2O_3$ matrices with ADP crystals grown in nanopores with a diameter of Ø25 nm

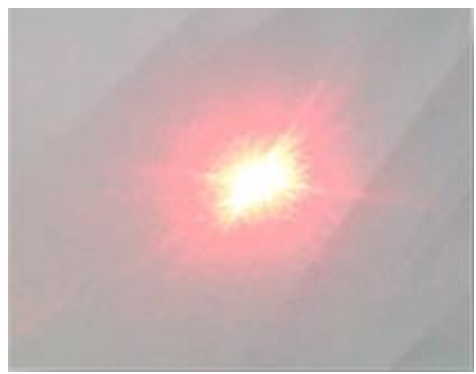

Fig. 7. Light scattering of a laser beam ($\lambda$=634nm) through the structure of $Al_2O_3$ without ADP crystals in the pores

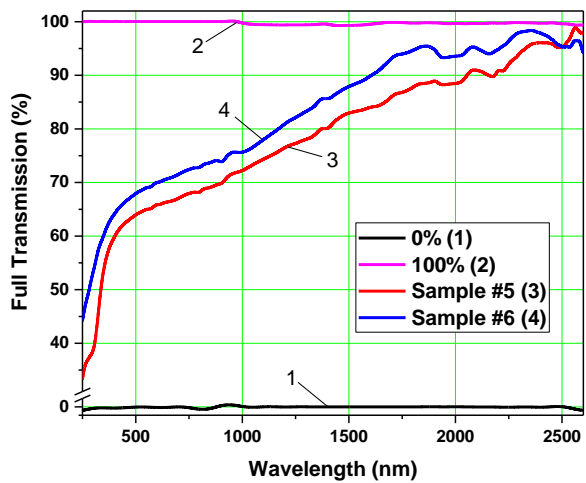

Fig. 8. Full (direct plus diffuse) transmission spectra of $Al_2O_3$ matrices with ADP crystals in the pores with the diameter of pores Ø 25 nm

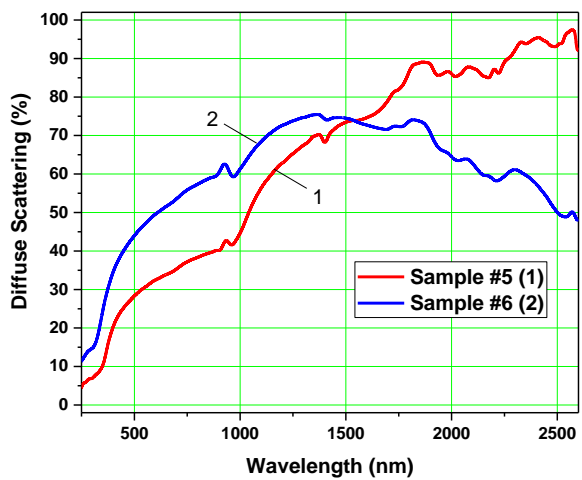

Fig. 9. Diffuse scattering spectra of samples #5 and #6 with ADP crystals with the diameter of pores Ø 25 nm

IV. CONCLUSIONS

Comparing the direct transmission spectra of $Al_2O_3$ matrices of 100 μm thickness and a diameter of pores 25 nm (Fig. 5) and the spectrum of transmission of $Al_2O_3$ samples with ADP crystals in the pores including the diffuse scattering (Fig. 8), it is evident that the change in transmission is due to diffuse scattering of light. From the analysis of results depicted in Fig. 9, it can be concluded that the diffuse scattering of sample #5 raises in the 0.6-3 μm wavelength range, while in sample #6 it decreases in the 1.5-3.0 μm wavelength range due to the nature of the transmitting behavior of samples #5 and #6 in the investigated region (Fig. 5). The reflection behavior in the investigated wavelength range of $Al_2O_3$ matrices filled with ADP crystals and pure $Al_2O_3$ matrices is due to the behavior of diffuse scattering in this investigated region of wavelengths (see Fig. 9). Thus, it can be concluded that the transmission and scattering of $Al_2O_3$ matrices with and without ADP crystals inside the nanopores depend to a large extent on the properties of the $Al_2O_3$ matrices and the technology of their production.

ACKNOWLEDGMENT

This result of investigation is a part of a project that has received funding from the European Union's Horizon 2020 research and innovation programme under the Marie Skłodowska-Curie grant agreement No 778156, as well as was supported by Ministry of Education and Science of Ukraine in the frames of the projects "Anisotropy" (# 0116U004136) and "Nanocomposite" (# 0116U004412).